\documentclass[12pt]{article}

\usepackage[dvips]{graphicx}
\usepackage{amsfonts}

\newcommand{\fourvector}{\mathbf}
\newcommand{\dd}{\mathrm{d}}

\newcommand{\g}{\fourvector{g}}
\newcommand{\M}{\mathcal{M}}
\newcommand{\diag}{\mathrm{diag}}
\newcommand{\K}{\fourvector{K}}
\newcommand{\Lied}[2]{\mathcal{L}_{#1}\,{#2}}
\newcommand{\h}{\fourvector{h}}
\newcommand{\N}{\mathcal{N}}
\newcommand{\on}[1]{|_{#1}}
\newcommand{\On}[1]{\Bigg|_{#1}}
\newcommand{\X}{\fourvector{X}}
\newcommand{\Y}{\fourvector{Y}}
\newcommand{\rr}{r_*}
\newcommand{\tc}{\frac{\partial\rr}{\partial r}}
\newcommand{\constant}{\mathrm{constant}}
\newcommand{\m}{\mathrm{m}}

\begin{document}
\title{\bf Global Structure of Certain Static Spacetimes (I)}
\author{Bin Zhou\thanks{E-mail: zhoub@bnu.edu.cn .} \\
  The Department of Physics, Beijing Normal University \\
  Beijing, 100875}
\date{April 14, 2001}
\maketitle

\begin{abstract}
  In this paper, static spacetimes with a topological structure of $\mathbb{R}
^2\times\N$ is studied, where $\N$ is an arbitrary manifold. Well known
Schwarzschild spacetime and Reissner-Nordstr\"om spacetime are special cases.
It is shown that the existence of a constant and positive surface gravity
$\kappa$ ensures the existence of the Killing horizon, with the cross section
homeomorphic to $\N$.
\end{abstract}

\section{Introduction}
\label{sectintro}

In general relativity, specific spacetimes often tell us much information
that is related to the properties of more generic spacetimes. The rather
simple spacetimes, if the trivial Minkowski spacetime is excluded,
should be the spherically symmetric static spacetimes: the sourceless
Schwarzschild spacetime, the Reissner-Nordstr\"om spacetime with
electromagnetism, and so on. Simple as they are, they have already possessed
variety of features which remain to be true for generic static, or, even
stationary, spacetimes. For example, if the stationary spacetime possesses
more structures such as additional symmetries or else, there is always the
Regge-Wheeler tortoise function encountered in these spacetimes and we find it
useful in various discussions: For the Kruskal extensions\cite{MTW,HawkEllis,
Wald,FrolovNovikov}, it is
the start point where the ingoing and outgoing Eddington-Finkelstein
coordinates and, thereafter, the Kruskal-Szekeres coordinates, are introduced;
In probing the thermodynamical properties of black holes, the tortoise function
is also widely used (see, for example, \cite{DamourRuffini,
Zhao}). In string theories, situations are similar.

Detailed discussion of specific spacetimes benefits us a lot. But restricting
our attention to particular spacetimes often sinks us into trouble: Sometimes
a conclusion is so directive to be derived that deep sense of the problem will
be ignored. It just like that you pave your way directly to your destination,
while wonderful scenery around have been missed because you are so quick to
reach where you want to go. The worst thing is that, since specific spacetimes
often possess additional structures such as the spherical symmetry (namely,
the $SO(3)$ symmetry)\cite{SachsWu,Ehlers}, compactness of the cross section
of the Killing horizon and so on, we don't know whether certain
conclusion is closely related to these properties. Example of this situation
may be the first law of black hole mechanics\cite{Wald,Townsend,BCH}. In the
ordinary cases, the entropy $S$ of the black hole is believed to be
proportional to the area of the cross section of the horizon. But, what if
the cross section of the Killing horizon is not compact, with the area being
infinite? The proof of the constancy of the surface gravity is another example
of depending on the additional symmetries of the spacetime.
To prove it, one has to prove that the surface gravity on the bifurcate sphere
is constant. For a sphere, this is certainly true. But, if the cross section
of the Killing horizon is not a sphere, or, if the Killing horizon has no
spherical symmetry, does the total proof remain valid?

In this paper we do not want to aim at any particular object. We just consider
static spacetimes of arbitrary dimension $d\geq 3$ which, topologically, is
the Cartesian product of $\mathbb{R}^2$ and an arbitrary manifold $\N$ endowed
with a Riemannian metric $\h_\N$. We don't assume any symmetry about the
Riemannian manifold $(\N,\h_\N)$. This, of course, has included Schwarzschild
spacetime and Reissner-Nordstr\"om spacetime as special examples. We try to
establish a somehow generic formulism for the extendibility of such spacetimes
and outline the extendibility of the spacetime as in the following:

The static spacetime admits the Killing vector field $\K$ as a globally defined
smooth vector field, which defines a globally smooth function $g_{tt}=
\g(\K,\K)$. The event horizon, if there is, consists of the zero points of
this function, implying that $\K$ is lightlike on it. The function $g_{tt}$ is
intrinsic, independent of the choice of any coordinates. There is also the
intrinsic smooth
function, the Regge-Wheeler tortoise function $\rr$. But it is, generally
speaking, not globally defined. The maximal domain of this smooth function is
called a region of the spacetime. If the closure of its maximal domain can't
cover the
spacetime, namely, not the whole spacetime itself, there must be another
Regge-Wheeler tortoise function whose maximal domain has no common subset with
the domain of $\rr$. If the closures  of these domains still can not cover
the whole spacetime, there must be a third tortoise function, and so on. Then
finally we have a set of tortoise functions as well as their corresponding
maximal domains (regions, as we call). Any pair of these regions have no common
subsets except for the empty set. Between a pair of adjacent regions there is
an event horizon. In this situation, each of such a region is called
extendible. There is the possibility that
the whole spacetime is extendible. If the surface gravity is a nonzero constant
on a horizon, and if the tortoise function tends to infinite as it approaches
the horizon, we can give the formulism showing how it can be extended, which
is quite a similar version of the Kruskal extension.

The paper is organized as the following. In \S\ref{sectstructure}, we state
the structure of the static spacetime whose extendibility would be considered.
In \S\ref{sectregion} we give the coordinate description to the region in
which the Killing vector field $\K$ is either timelike or spacelike.
Two special lightlike vector fields are given for future use.
In \S\ref{sectgeodesic} we investigate the completeness of the integral curves
of the above two lightlike vector fields so that we can trace the behavior of
the tortoise function as it approaches the zero points of $g_{tt}$.
In \S\ref{sectKruskal} we derive the formulism as how the region can be
extend out of the horizon. And \S\ref{sectdiscussion} is the closure part of
this paper, the discussion and conclusions.

\section{The Structure of Spacetimes Considered Here}
\label{sectstructure} 

In this paper, the spacetime we considered is a connected static spacetime
$(\M,\g)$
of $d \geq 3$ dimensional, where $\g$ is the metric tensor of the
spacetime with the signature of $\diag(-1,1,\ldots,1)$.
The Killing vector field is denoted by $\K$. Hence $\K$ satisfies two
equations, the Killing equation
\begin{equation}
  \Lied{\K}{\g} = 0
\label{Lieeq4K}
\end{equation}
and the Fr\"obenius condition
\begin{equation}
  \tilde{\K}\wedge\dd\tilde{\K} = 0.
\label{Frobenius4K}
\end{equation}
Note that, in the above equation, $\tilde{\K}$ is the 1-form resulted by
raising the index of $\K$ with $\g$: In the language of abstract
indices\cite{Wald,Penrose}, $\K$ is denoted by $K^a$, and $\tilde{\K}$ is
denoted by $K_a = g_{ab} K^b$.

To simplify our problem, we shall assume further that the spacetime $\M$ is
a Cartesian product of $\mathbb{R}^2$ and $\N$, where $\N$ is an arbitrary
$d-2$ dimensional manifold endowed with a Riemannian metric $\h_\N$. We assume
that the Killing vector field $\K$ is always ``tangent" to the directions of
$\mathbb{R}^2$. That is, let $\pi:\M\longrightarrow\N$ be the projection from
$\M = \mathbb{R}^2\times\N$ to $\N$, then $\pi_*(\K\on{p})=0$ at any point
$p$ in $\M$. If we define the symmetric tensor field $\h$ on $\M$ to be
\begin{equation}
  \h = \pi^*\h_\N,
\label{hdef}
\end{equation}
the Killing vector field $\K$ will be orthogonal to the symmetric tensor
field $\h$ on $\M$. That is, if the 1-forms $h_{ab}K^b$ and $K^bH_{ba}$ are
denoted by $\h\cdot\K$ and $\K\cdot\h$, respectively, we could claim that
\begin{equation}
  \h\cdot\K = \K\cdot\h = 0.
\label{Korth2h}
\end{equation}

In order to make the calculations easier, we may introduce local coordinates
$(t,r,x^i)$ ($i=1,\ldots,d-2$) into certain region of $\M$, with $t$ and
$r$ along the two
directions of $\mathbb{R}^2$ and $x^i=\pi^* x_\N^i$ being the pull back of
local coordinate $x^i_\N$ on $\N$. Now we
can write $\K$ and $\g$ as
\begin{displaymath}
  \K = K_t\,\frac{\partial}{\partial t} + K_r\,\frac{\partial}{\partial r}
\end{displaymath}
and
\begin{equation}
  \g = g_{tt}\,\dd t \otimes \dd t
  + g_{tr}\,\big(\dd t\otimes\dd r + \dd r\otimes \dd t\big)
  + g_{rr}\,\dd r\otimes\dd r 
  + \fourvector{\beta}\otimes\dd r + \dd r\otimes\fourvector{\beta}
  + C\,\h,
\label{metricg}
\end{equation}
where $\fourvector{\beta} = B_i\, \dd x^i$ is a differential 1-form on $\M$.
According to Eqs.(\ref{hdef}) and (\ref{Korth2h}), $\h$ can
be expressed as
\begin{displaymath}
  \h = h_{ij}\,\dd x^i\otimes\dd x^j,
\end{displaymath}
where the functions $h_{ij}$'s are independent of the coordinates $t$ and $r$.
Obviously, the Lie derivative of $\h$ with respect to $\K$ is zero.

In the neighborhood of a point at which $\K$ is nonzero, we can always choose
its neighborhood properly so that $\K$ is nonzero everywhere in it. When the
coordinate system $(t,r,x^i)$ is defined in this neighborhood, we can set $t$
and $r$ carefully chosen in order that $K_t=1$, $K_r=0$ in the above
equation:
$\K=\frac{\partial}{\partial t}$.
Since the spacetime is static, coordinates can be chosen such that either
$g_{tt}$ or $g_{tr}$ is zero.
In this paper, our topic is mainly focused on regions where $\K$ is nonzero
everywhere.

\section{In the Region Where $\K$ Is Timelike or Spacelike}
\label{sectregion}

First let us consider the coordinate neighborhood in which $\K$ is either
timelike or spacelike. In such a neighborhood, the coordinates $t$ and $r$
can be organized so that $g_{tr}=g_{rt}=0$.
Then Eq.(\ref{Lieeq4K}) implies that the functions $g_{tt}$, $g_{rr}$,
$g_{tr}=g_{rt}$, $B_i$ and $C$ are all independent of $t$.

In this case, a coordinate transformation $(t,r,x^i)\longrightarrow
(t,\rr,x^i)$ can be used for simplifying the discussion. Here
\begin{equation}
  \rr = \rr(r,x)
\label{tortoise}
\end{equation}
is a function that will turn the metric $\g$ into the form of
\begin{equation}
  \g = g_{tt}\,\big(\dd t\otimes\dd t - \dd\rr\otimes\dd\rr\big)
  + \fourvector{\beta_*}\otimes\dd\rr + \dd\rr\otimes\fourvector{\beta_*}
  + \h_*.
\label{newformg}
\end{equation}
A bit of calculation reveals that
\begin{eqnarray}
  \tc & = & \sqrt{-\frac{g_{rr}}{g_{tt}}},
\label{tcdef}
\\
  \fourvector{\beta_*} & = & \sqrt{-\frac{g_{tt}}{g_{rr}}}\fourvector{\beta}
  + g_{tt}\frac{\partial\rr}{\partial x^i}\dd x^i,
\label{betastar}
\\
  \h_* & = & C\,\h - g_{tt}\frac{\partial\rr}{\partial x^i}
  \frac{\partial\rr}{\partial x^j}\dd x^i\otimes\dd x^j
  - \frac{\partial\rr}{\partial x^i}\sqrt{-\frac{g_{tt}}{g_{rr}}}\,
  \big(\fourvector{\beta}\otimes\dd x^i + \dd x^i\otimes\fourvector{\beta}
  \big).
\end{eqnarray}
It is easy to see that $\h_*$ is the induced metric of $\g$ on the surface
determined by $t=\constant$ and $\rr=\constant$.
When we turn back to the well known cases such as the Reissner-Nordstr\"om
spacetime, the function $\rr$ is just the Regge-Wheeler tortoise coordinate
function. Since the Lie derivatives of $\fourvector{\beta}$ and $\h$ with
respect to $\K$ are zero, we can see that
\begin{equation}
  \Lied{\K}{\fourvector{\beta_*}} = 0, \qquad
  \Lied{\K}{\h_*} = 0
\label{Liedtortoise}
\end{equation}
with a peer and, still,
$\fourvector{\beta_*}(\X) = 0$, $\h_*(\X,\Y) = 0$ whenever $\X$ satisfies
$\pi_*\X=0$.

Now we want to find out that whether there are vector fields $\X_+$ and
$\X_-$ on $\M$ satisfying:
\begin{itemize}
\item[(1)] They lie in the $\mathbb{R}^2$ plains, namely, $\pi_*\X_\pm = 0$;
\item[(2)] They are lightlike vector fields, namely, $\g(\X_+,\X_+) =
  \g(\X_-,\X_-) = 0$;
\item[(3)] They are geodesic vector fields, namely, $\nabla_{\X_+}\X_+ =0$
  and $\nabla_{\X_-}\X_-=0$.
\end{itemize}
Using the coordinate $(t,\rr,x^i)$, we can determine these vector fields, if
there exists, to be
\begin{displaymath}
  \X_\pm = X\,\bigg(\frac{\partial}{\partial t}\pm\frac{\partial}{\partial\rr}
  \bigg),
\end{displaymath}
according to the first two conditions. It is not necessary to calculate the
Christoffel symbols for $\nabla_{\X_\pm}\X_{\pm}$. In fact, there is the
formula\footnote{
Due to the well known formula, in the language of abstract indices,
$\Lied{\X}g_{ab}=\nabla_a X_b+\nabla_bX_a$.
} $\X_+\cdot\Lied{\X_+}\g = \nabla_{\X_+}\tilde{\X}_+
  + (\nabla\tilde{\X}_+)\cdot\X_+
=\nabla_{X_+}\tilde{\X}_+$, and we know that the Lie derivatives are more
easier to calculate. It is easy to verify that
\begin{displaymath}
  \nabla_{\X_\pm}\tilde{\X}_\pm = X\,\bigg[\,\frac{\partial}{\partial t}
  \big(g_{tt}X\big) \pm\frac{\partial}{\partial\rr}\big(g_{tt}X\big)\bigg]
  \big(\dd t\mp\dd\rr\big)
  \pm \Lied{\X_\pm}{\big(X\fourvector{\beta_*}\big)}
\end{displaymath}
We can set $X = \frac{1}{|g_{tt}|}$, yielding
\begin{eqnarray}
  & & \X_{\pm} = \frac{1}{|g_{tt}|}\bigg(
  \frac{\partial}{\partial t} \pm \frac{\partial}{\partial\rr}\bigg),
\\
  & & \nabla_{\X_\pm}\tilde{\X}_\pm = \pm \Lied{\X_\pm}{\bigg(
  \frac{1}{|g_{tt}|} \fourvector{\beta_*}\bigg)}
  = \pm i_{\X_\pm}\dd\bigg(\frac{1}{|g_{tt}|} \fourvector{\beta_*}\bigg).
\end{eqnarray}
Notice that the function $\rr$ is not uniquely determined by Eq.(\ref{tcdef}).
For a known function $\rr = \rr(r,x)$ satisfying this equation, one can always
give another function $\rr^\prime = \rr - \psi(x)$ where $\psi(x)$ is an
arbitrary function of $x^i$'s only. If we define $\fourvector{\beta}_*^\prime$
for $\rr^\prime$ as that in Eq.(\ref{betastar}) for $\rr$, then
\begin{equation}
  \dd\psi = \frac{1}{|g_{tt}|}\big(\fourvector{\beta}_*
  - \fourvector{\beta}_*^\prime\big)
\end{equation}
is a locally exact 1-form. If $\frac{1}{|g_{tt}|}\fourvector{\beta}_*$ itself
is exact,
we can choose $\psi(x)$ properly so that $\fourvector{\beta}_*^\prime=0$.
This makes the 1-form $\frac{1}{|g_{tt}|}\fourvector{\beta}_*$
significant. In fact, it needs not necessarily to be exact to let the vectors
$\X_\pm$ be geodesic, for example, $\frac{1}{|g_{tt}|}\fourvector{\beta}_*$
can be just closed, or else, $\frac{1}{|g_{tt}|}\fourvector{\beta}_*=B_{*\,i}\,
\dd x^i$ with the coefficients $B_{*\,i}$ independent of $t$ and $\rr$.

In the following, we will assume that $\frac{1}{|g_{tt}|}\fourvector{\beta}_*$
is an exact 1-form. As discussed in the above paragraph, it is equivalent to
assume that $\fourvector{\beta}_*=0$. This has covered the well known
Schwarzschild spacetime and Reissner-Nordstr\"om spacetime as special cases:
In the case of Reissner-Nordstr\"om spacetimes, the Riemannian manifold $\N$
is a 2-dimensional sphere. The metric components $g_{tt}$ and $g_{rr}$ are
independent of $(x^1,x^2)=(\theta,\varphi)$ with the 1-form
$\fourvector{\beta}$ being zero, making the Regge-Wheeler tortoise coordinate
function $\rr$ to depend only on $r$ and $\fourvector{\beta}_*=0$. In the
following discussion, we do not need $g_{tt}$ and $g_{rr}$ to be independent of
$x^i$, and we do not need $\fourvector{\beta}$ to be zero, either. What we
need is just that a function $\rr=\rr(r,x)$ can be picked making the 1-form
$\fourvector{\beta}_*$ to be zero.

Immediately, the 1-forms
\begin{equation}
  \tilde{\X}_+ = \eta_{tt}\,\dd u, \qquad \tilde{\X}_- = \eta_{tt}\,\dd v
\label{tildeX}
\end{equation}
are (at least locally in the neighborhood where the coordinates $t$, $r$
and $x^i$ are defined) exact, where $\eta_{tt}$ is the sign of $g_{tt}$, and
$u=t-\rr$ and $v=t+\rr$ are the outgoing and ingoing lightlike coordinate
functions,
respectively. In the coordinate neighborhood, the component form of the metric
$\g$ in Eq.(\ref{metricg}) is well defined. The question is, what would happen
when it tends to the boundary of the coordinate neighborhood.

If the symmetric tensor field $\h_*$ tends to be degenerate or undefinable at
the boundary of the coordinate neighborhood, we can believe that the spacetime
could not extend out of the boundary any more. If, however, only the function
$g_{tt}$ tends to be zero or infinity at the boundary, the conclusion may be
quite
different. What it will be depends on the coordinate free expressions. This is
why we want to introduce the two lightlike geodesic vector fields $\X_+$ and
$\X_-$.

\section{The Geodesic Curves of $\X_\pm$}
\label{sectgeodesic}

As we have seen, there are the relations
\begin{equation}
  \X_+u = \dd u\,(\X_+) = \eta_{tt}\,\tilde{\X}_+(\X_+)
  = \eta_{tt}\,\g(\X_+,\X_+) = 0, \qquad
  \X_-v = \eta_{tt}\,\tilde{\X}_-(\X_-) = 0
\end{equation}
and
\begin{equation}
  \X_+v = \dd v(\X_+) = \eta_{tt}\,\tilde{\X}_-(\X_+) = \frac{2}{|g_{tt}|},
  \qquad \X_-u = \frac{2}{|g_{tt}|}. 
\end{equation}
The meaning of these relations can be interpreted as in the following. Suppose
$\gamma_+:I\longrightarrow\M$, $\lambda\longmapsto\gamma_+(\lambda)$ is an
integral curve of the lightlike vector $\X_+$, where $I$ is some interval
of $\mathbb{R}$ containing 0 and $\gamma_+(0)=p$ lies in the
coordinate neighborhood. Obviously $\gamma_+$ is a geodesic lightlike curve
which can be described, in the coordinate language, as
\begin{displaymath}
  u(\lambda) = u = \constant, \ v(\lambda)=v(\gamma_+(\lambda)),
  \ x^i(\lambda)= x^i = \constant.
\end{displaymath}
On the one hand, the tangent vector $\dot{\gamma}_+(\lambda)$ can be written
in the coordinate system $(u,v,x^i)$ as
\begin{displaymath}
  \dot{\gamma}_+(\lambda) = \frac{\dd v}{\dd\lambda}
  \frac{\partial}{\partial v}\On{\gamma_+(\lambda)}
  = \frac{1}{2}\frac{\dd v}{\dd\lambda}\bigg(
  \frac{\partial}{\partial t}+\frac{\partial}{\partial\rr}\bigg)
  = \frac{|g_{tt}|}{2}\frac{\dd v}{\dd\lambda}\,\X_+\On{\gamma_+(\lambda)}.
\end{displaymath}
On the other hand, $\dot{\gamma}_+(\lambda)$ is nothing else but
$\X_+\On{\gamma_+(\lambda)}$. Hence we obtain
\begin{equation}
  \frac{\dd v}{\dd\lambda} = \frac{2}{|g_{tt}(\frac{v(\lambda)-u}{2},x)|}.
\label{functionv}
\end{equation}
Combined with $\frac{\dd u}{\dd\lambda}=0$, it reveals that
\begin{equation}
  \frac{\dd t}{\dd\lambda} = \frac{\dd\rr}{\dd\lambda}
  = \frac{1}{|g_{tt}(\frac{v(\lambda)-u}{2},x)|}.
\label{tandr}
\end{equation}
For a pair of affine parameters $\lambda_0$ and $\lambda_1$, let $v_0
=v(\lambda_0)$ and $v_1=v(\lambda_1)$, then
\begin{eqnarray}
  v_1-v_0 & = & \int^{\lambda_1}_{\lambda_0}\frac{2}{|g_{tt}(\rr(\lambda),x)|}
  \,\dd\lambda,
\label{intv} \\
  \lambda_1-\lambda_0 & = & \frac{1}{2}\int^{v_1}_{v_0}
  |g_{tt}(\frac{v-u}{2},x)|\,\dd v
  = \int^{\frac{v_1-u}{2}}_{\frac{v_0-u}{2}}|g_{tt}(\rr,x)|\,\dd\rr.
\end{eqnarray}

Similarly, for the integral curve $\gamma_-: I \longrightarrow\M,\lambda
\longrightarrow\gamma_-(\lambda)$ which is described by
\begin{displaymath}
  u(\lambda) = u(\gamma_-(\lambda)),\ v(\lambda)=v=\constant,\ 
  x^i(\lambda)=x^i=\constant,
\end{displaymath}
we can obtain
\begin{equation}
  \frac{\dd u}{\dd\lambda} = \frac{2}{|g_{tt}(\frac{v-u(\lambda)}{2},x)|}
\label{functionu}
\end{equation}
and, for a pair of affine parameters $\lambda_0$ and $\lambda_1$ with $u_0
=u(\lambda_0)$ and $u_1=u(\lambda_1)$,
\begin{eqnarray}
  u_1-u_0 & = & \int^{\lambda_1}_{\lambda_0}
  \frac{2}{|g_{tt}(\frac{v-u(\lambda)}{2},x)|}\,\dd\lambda, \\
  \lambda_1-\lambda_0 & = & \frac{1}{2}\int^{u_1}_{u_0}|g_{tt}
  (\frac{v-u}{2},c)| \,\dd u 
  = \int^{\frac{v-u_0}{2}}_{\frac{v-u_1}{2}}|g_{tt}(\rr,x)|\,\dd\rr.
\end{eqnarray}

These two sets of curves are closely related to each other. Let $v=S(\lambda,
u,x)$ be a solution of the differential equation (\ref{functionv}), then it is
very easy to verify that
\begin{equation}
  u(\lambda)=2v-S(-\lambda,v,x)
\label{u2v}
\end{equation}
is a solution to the equation (\ref{functionu}). Both $v(\lambda)$ and
$u(\lambda)$ are increasing functions of $\lambda$.

If the function $g_{tt}$ tends to zero along one of the curves when the
affine parameter $\lambda$ increases (or decreases), we may ask whether
the parameter $\lambda$ is finite or not before it reaches the zero point
of $g_{tt}$. If it is, we call that the geodesic is incomplete in that
direction. Otherwise we say it is complete in that direction. Take the
direction in which $\lambda$ increases, for example. If the geodesic is
complete in that direction, then $\lambda$ tends to $+\infty$ as it approaches
the ``zero point" of $g_{tt}$. Then Eq.(\ref{tandr}) indicates that this
``point",
if it exists in the spacetime, corresponds to $\rr=+\infty$. We just call
na\"\i vely that the ``zero point" of $g_{tt}$ ``is located at" where
$\rr=+\infty$.

As for incomplete geodesics, cases will be complicated. Suppose that $g_{tt}$
tends to zero along the curve of $\X_+$ as $\lambda$ tends to a finite value
$\lambda_\m$. If the velocity at which $g_{tt}$ tends to zero is not slower
than that of $\frac{1}{\lambda-\lambda_\m}$, the ``zero point" of $g_{tt}$
is still located at where $\rr=+\infty$, or accurately speaking, at where
$v=+\infty$ (cf Eqs.(\ref{tandr}) and
(\ref{intv})). And, because the relationship (\ref{u2v}) of the integral
curves of $\X_+$ and $\X_-$, we can safely assert that one the geodesic of
$\X_-$ will be incomplete as the parameter increases, and that the ``zero
point" of $g_{tt}$ is located at where $u=-\infty$, and \textit{vice versa}.
As we have known, ``locations" where $u=-\infty$ and where $v=+\infty$ are
not the same. Both of these locations give $\rr=+\infty$. The most miserable
thing is that the geodesics are incomplete and leading $g_{tt}$ tends to
zero slower than or as quickly as $\frac{1}{\lambda-\lambda_\m}$, which
locates the ``zero points" of $g_{tt}$ at where $\rr=\rho(x)$ for some
function $\rho(x)$. Topic like this will be left to proceeding
paper. In this paper, we discuss cases in which the ``zero points" of $g_{tt}$
are located at where $\rr=\infty$ only.

The Schwarzschild spacetime and the Reissner-Nordstr\"om spacetime match the
above discuss. The latter has regions in which every geodesic is incomplete
in both directions, giving the location of $\rr=+\infty$ as well as $\rr
= -\infty$. As in these examples, generally speaking,
the spacetime can be extended far away out of the coordinate
neighborhood. In the following, we shall concentrate our attention to discuss
it. 

But before we go ahead, we had better give a summary to the above discussion. 

Notice that the
metric tensor $\g = g_{tt}\,(\dd t\otimes\dd t - \dd\rr\otimes\dd\rr) + \h_*$
is rather simple, where we have assumed that $\h_*$ behaves very well, no
matter in or out of the coordinate neighborhood. And $g_{tt}$ may encounter
some zero points on the boundary of this neighborhood. Experience in
studying Schwarzschild spacetime or Reissner-Nordtr\"om spacetime implies that
these zero points are often not the singularities of the whole spacetime. They
are often believed to be due to the choice of the coordinates. It is somewhat
right but not the case, in fact. Why? Because the function $g_{tt}$ has its own
significance as the square norm of the Killing vector field $\K$, $\g(\K,\K)$.
In this sense it stands there independent of how the coordinates have been
chosen. Thus, to treat its zero points as the illness of the coordinates, as
that is misunderstood by quite a large amount of people, is unfair for the
coordinates. Zero points of $g_{tt}$, were they not the singularities, are
really special points of the spacetime that must be considered especially.

\section{The Kruskal Extension}
\label{sectKruskal}

For the metric tensor $\g$ with the ``zero points" of $g_{tt}$ located at
where $\rr=-\infty$, or equivalently, $u=+\infty$ and $v=-\infty$, we
imitate what has been done in the Kruskal spacetime\cite{Krusk},
introducing two functions
\begin{equation}
  U = -e^{-\kappa u}, \qquad V=e^{\kappa v},
\label{UVdef}
\end{equation}
where $\kappa$ is a positive constant to be determined. They are increasing
functions of $u$ and $v$, respectively. And the ``surfaces" $u=+\infty$ and
$v=-\infty$ are described as $U=0$ and $V=0$, respectively. Therefore the
coordinate neighborhood of system $(t,r,x^i)$ is contained in the region
$U<0$ and $V>0$. Using these two variables, the metric $\g$ looks like
\begin{equation}
  \g = G(U,V,x)\,\big(\dd U\otimes\dd V + \dd V\otimes\dd U\big) + \h_*,
\label{ginUV}
\end{equation}
where
\begin{equation}
  G(U,V,x) = \frac{1}{2\kappa^2}\,g_{tt}\,e^{-2\kappa\rr}
\label{Gdef}
\end{equation}
is a function of $U$, $V$ and $x^i$. It is well known that the function $\rr$
can be treated as a function of $U$ and $V$, defined by
$\rr=\frac{1}{2\kappa}\ln(-UV)$.

If there is a positive constant $\kappa$ such that
\begin{equation}
0< G_0 = \lim_{\rr\to-\infty}|g_{tt}(\rr,x)|\,e^{-2\kappa\rr} <+\infty,
\label{G0def}
\end{equation}
then $\displaystyle{\lim_{u\to+\infty}}G(u,v,x)=\displaystyle{
\lim_{v\to-\infty}}G(u,v,x) = \frac{\eta_{tt}}{2\kappa^2}G_0$ is a nonzero
function of $x^i$'s, and we
are able to embed isotropically the coordinate neighborhood of system
$(t,r,x^i)$ into a region that contains points where $U=0$ or $V=0$. That
is, the (or some of) zero points of $g_{tt}$ are not singularities. Instead,
they are points of the spacetime, or else, the spacetime can be extended to
include them if they were not parts of the spacetime originally. The constant
$\kappa$ can be calculated, if it exists. In fact, L'Hospital's rule can be
applied to Eq.(\ref{G0def}) since $g_{tt}$ tends to zero as we have assumed.
Thus $G_0 = \displaystyle{\lim_{\rr\to-\infty}}\frac{\eta_{tt}}
{2\kappa\,e^{2\kappa\rr}} \frac{\partial g_{tt}}{\partial\rr}
=G_0\,\displaystyle{\lim_{\rr\to-\infty}}\frac{1}{2\kappa\,g_{tt}}
\frac{\partial g_{tt}}{\partial\rr}
= \displaystyle{\frac{G_0}{2\kappa}}\lim_{\rr\to-\infty}\frac{\partial}
{\partial\rr}\ln g_{tt} \neq 0$, from which we obtain the positive constant
\begin{equation}
  \kappa = \frac{1}{2}\lim_{\rr\to-\infty}\frac{\partial}{\partial\rr}
  \ln|g_{tt}|
  = \lim_{r\to r_\m}\frac{1}
  {\sqrt{|g_{rr}|}}\,\frac{\partial\sqrt{|g_{tt}|}}{\partial r},
\label{kappa}
\end{equation}
where $r\to r_\m$ corresponds to $\rr\to-\infty$. The above formulas have been
derived and widely used in references \cite{Zhao}.
One can verify that, when applied to the Schwarzschild spacetime or
the Reissner-Nordstr\"om spacetime, they give the resulted $\kappa$ as the
surface gravity. Intuitively, they reveal the demand that $g_{tt}$ must behave
much like $e^{\kappa(v-u)} = e^{2\kappa\rr}$ nearby the zero points of
$g_{tt}$. Notice that $g_{tt}$ and $g_{rr}$ often depend on the coordinates
$x^i$. But, the limit value $\kappa$ in the above equation must be independent
of any coordinates. This has imposed strong limitation upon the metric $\g$.

In the regular region, namely, where $U\neq 0$ and $V\neq 0$, the partial
derivatives of $G$ with respect to $U$ and $V$ are smooth. For example,
\begin{equation}
  \frac{\partial G}{\partial U} = \frac{1}{4\kappa^2U}\frac{\partial g_{tt}}
  {\partial\rr}e^{-2\kappa\rr}
  - \frac{1}{2\kappa^2U}g_{tt}e^{-2\kappa\rr}.
\end{equation}
As the variable $U$ tends to $0^-$, the limit of the above is of type $\frac{0}
{0}$ for a nonzero $V$. Since the partial derivative of $G$ with respect to
$U$ at $U=0$ is
\begin{displaymath}
  \frac{\partial G}{\partial U}\On{U=0^-} = \lim_{U\to 0^-}\frac{
  G(U,V,x) - G(0,V,x)}{U}
  = \lim_{U\to 0^-}\frac{\partial G}{\partial U}
\end{displaymath}
if the last limit exists, and hence it will be continuous at $U=0$. If we write
down the partial derivative of $G$ with respect to $V$, the expression
indicates that it is also continuous at $V=0$, implying that both the
partial derivatives are continuous on the whole region of $U\leq 0$ and
$V\geq 0$. If all the limits of various partial derivatives of $G$ exist
as $U\to 0^-$ and $V\to 0^+$, the metric can be smoothly extended to the
region which includes the zero points of $g_{tt}$, as what we have known
to the Kruskal spacetimes.

Suppose that there is another spacetime (or region, we may call it the region
II and call the spacetime region discussed in the above as the region I
for convenience) that is similar to the above discussed: the region II is
topologically also a Cartesian product of $\mathbb{R}^2$ and $\N$, endowed
with a metric $\g^\prime$ admitting a Killing vector field $\K^\prime$ with
similar properties. Suppose that the zero points of $g^\prime_{tt}$ are
located at where $\rr^\prime=+\infty$ this time, and the positive constant
$\kappa$ is the same as the above, enabling
\begin{equation}
  U^\prime = e^{\kappa u^\prime}, \qquad V^\prime = - e^{-\kappa v^\prime}
\end{equation}
such that
\begin{equation}
 \g^\prime = G^\prime(U^\prime,V^\prime)(\dd U^\prime\otimes\dd V^\prime
  + \dd V^\prime\otimes\dd U^\prime)+\h^\prime_*
\end{equation}
can be smoothly extended to include the points where $U^\prime=0$ as well as
$V^\prime=0$. If, at $U^\prime=0$ and $V^\prime=0$, respectively, all the
partial derivatives of $G^\prime$ of various
order with respect to $U^\prime$ and $V^\prime$ equal to the corresponding
ones of $G$ at $U=0$ and $V=0$, respectively, the region I and the region
II can be glued together by identifying $(U=0,V,x)$ with $(U^\prime=0,V^\prime,
x)$ provided that the remainder part $\h_*$ and $\h^\prime_*$ can be fixed
together smoothly.
And, we know that we can copy the region I to be a new region I$^\prime$
with the variables $U$ and $V$ of opposite values of those in the region I,
and the region II can also be copied to be a new region II$^\prime$ in the
same way. Thus region I and region II$^\prime$ can be glued together by
identifying $(U,V=0,x)$ with $(U^\prime,V^\prime=0,x)$. In the same way,
regions I$^\prime$ and II$^\prime$, regions I$^\prime$ and II can also be
glued together, respectively. In this way, we have recovered a Kruskal
spacetime, just as that has been done for the well known examples. 

The expression of metric $\g$ in Eq.(\ref{ginUV}) indicates that hypersurfaces
$U=\constant$ or $V=\constant$ are lighlike. The lightlike hypersurface $U=0$
with $V\neq 0$ or $V=0$ with $U\neq 0$, the event horizons, consist of zero
points of $g_{tt}=\g(\K,\K)$.  That is, the Killing
vector field $\K$ becomes lighlike on these hypersurfaces while it is either
timelike or spacelike off them. Using the coordinates $U$, $V$ and $x^i$,
the Killing vector field reads
\begin{equation}
  \K = \kappa\, V\frac{\partial}{\partial V}
  - \kappa\, U\frac{\partial}{\partial U}.
\label{KinUV}
\end{equation}
Obviously, on the horizon $U=0$, $\K=\kappa V\frac{\partial}{\partial V}$ is
tangent to it. And it is for the same reason that
$\K=-\kappa U \frac{\partial}{\partial U}$ is also tangent to the horizon
$V=0$. Hence the event horizons are Killing horizons, for, on these horizons,
the 1-form $\tilde{\K}$ is still well defined. To see it, we begin from the
off horizon expression
\begin{displaymath}
  \tilde{\K} = g_{tt}\,\dd t = \frac{g_{tt}}{2\kappa}\bigg(
  \frac{1}{V}\dd V - \frac{1}{U}\dd U\bigg).
\end{displaymath}
Using the function $G$, which is well defined in the neighborhoods of the
horizons, the above expression becomes
\begin{equation}
  \tilde{\K} = \kappa\,G\,\big(\,V\dd U - U\dd V),
\end{equation}
coinciding with Eq.(\ref{KinUV}). In order to calculate $\nabla_{\K}\K$ on the
horizons, we first extend $\K\on{U=0}$ or $\K\on{V=0}$ to be a vector field
defined in the neighborhood of the horizon. The extension varies arbitrarily.
We simply choose $\K$ as that. Eq.(\ref{Lieeq4K}), or equivalently, $\nabla_a
K_b + \nabla_b K_a = 0$ in the language of abstract indices, gives the formula
$\K\cdot\Lied{\K}\g = 0 = \nabla_\K\tilde{\K} + \frac{1}{2}\dd g_{tt}$.
Hence
\begin{equation}
  \nabla_\K\tilde{\K} = - \frac{1}{2}\dd g_{tt}
  = \frac{\kappa\, G}{2g_{tt}}\,\frac{\partial g_{tt}}{\partial\rr}
  \big(\,U\dd V + V\dd U\big) 
  - \frac{1}{2}\,\frac{\partial g_{tt}}{\partial x^i}\dd x^i.
\end{equation}
Applying Eq.(\ref{kappa}) here and noticing that $g_{tt}\on{U=0}=0$, we
obtain that
\begin{equation}
  \nabla_{\K}\tilde{\K}\On{U=0} = \kappa^2 GV\,\dd U
  = \kappa\,\tilde{\K}\On{U=0}, \qquad
  \nabla_{\K}\tilde{\K}\On{V=0} = \kappa^2 GU\,\dd V
  = -\kappa\,\tilde{\K}\On{V=0}.
\end{equation}
Thus it is confirmed that $\kappa$ \emph{is} the surface gravity.

Note that points where $U=V=0$ are also admitted in the spacetime in this
case. Such points are contained in a submanifold of $\M$, which is homeomorphic
to the manifold $\N$. They are where the Killing vector field $\K$ vanishes
and where the Killing horizons extend out, hence the submanifold is the
bifurcate manifold, not necessarily a sphere because we didn't assume that
for $\N$.

\section{Discussion and Conclusions}
\label{sectdiscussion}

For a stationary spacetime $(\M,\g)$ that admitting the Killing vector field
$\K$, the function $g_{tt}=\g(\K,\K)$ is a globally defined intrinsic function
which may have some zero points as the special points of the spacetime. For
the spacetimes as we have discussed in this paper, the Regge-Wheeler tortoise
function $\rr$ is also intrinsic except that it is, generally speaking, not
globally defined. The maximal domain in which it can be defined smoothly is a
region that is often called the exterior region, the black hole region or the
white hole region, and so on. If the tortoise function $\rr$ can not be
extended smoothly to cover the whole spacetime, there must be another tortoise
function whose domain can't be enlarged to intersect with that of $\rr$. Hence
the spacetime possesses regions, each with a Regge-Wheeler tortoise function
defined smoothly in it and not able to be extended smoothly anymore. Since
the spacetime we considered is connected, these regions must be separated
by event horizons, on which the Killing vector field is lighlike, or,
equivalently, the globally defined function $g_{tt}$ vanishes on them. Now that
spacetime is static, these event horizons are Killing horizons, with the
surface gravity being constant and nonzero.

I. R\'acz and R. M. Wald\cite{RaczWald} 
have studied a globally hyperbolic stationary spacetime containing a black
hole but not a white hole, with the event horizon of the black hole being a 
Killing horizon with compact cross sections. They proved that, if the surface
gravity is nonzero and constant throughout the Killing horizon, the spacetime
can be globally extended such that the (image of the) horizon is a proper
subset of a regular bifurcate Killing horizon in the enlarged spacetime.
The spacetime we considered in this paper is, of course, different from what
is studied by I. R\'acz and R. M. Wald, but we should be aware of the
similarity between the results. Both the cases have implied that the constancy
and nonzero property of the surface gravity are closely related to the
extendibility of the spacetime. For the extremal Reissner-Nordstr\"om
spacetime, for example, the surface gravity is zero everywhere on the horizon,
the spacetime regions could not be extended as what had been done with the
non-extremal case.

Of course, we have not fulfilled the tasks which were put forward in the
beginning of this paper. However, the narrowered
study has covered most of the well known examples of static spacetimes, from
which we find that the horizons, homeomorphic to the manifold $\N$, can be of
arbitrary shape, including those with the volume not finite.

\vskip 20mm

\begin{center}
  \textbf{\large Acknowledgements}
\end{center}
  The author is grateful to Professor Z. Zhao for his urgement. Some of the
contents have been discussed with Professor Y. K. Lau and Dr. X. N. Wu long
before this paper is prepared, and the author wants to thank them for their
criticisms.

\end{document}